\documentclass[12pt,preprint]{aastex}
\begin{document}
\title{INDEPENDENT MEASUREMENTS OF THE DYNAMICAL MASSES OF SIX GALAXY CLUSTERS IN THE LOCAL UNIVERSE}
\author{Jounghun Lee}
\affil{Astronomy Program, Department of Physics and Astronomy, Seoul National University, 
Seoul 08826, Republic of Korea} 
\email{jounghun@astro.snu.ac.kr}
\begin{abstract}
We present independent measurements of the masses of the galaxy clusters in the local universe by employing the Dynamical Mass Estimator 
(DME) originally developed by Falco et al in 2014.  In the catalog of the galaxy groups/clusters constructed by Tempel et al. from the Sloan 
Digital Sky Survey Data Release 10, we search for those as the targets around which the neighbor galaxies constitute thin straight filamentary 
structures in the configuration space spanned by the redshifts and the projected distances.   Out of the $29$ Sloan clusters that have 
$100$ or more member galaxies, a total of six targets are found to have filamentary structures in their bound zones. 
For each of the six targets, we construct the profile of the recession velocities of the filament galaxies, which depends on the cluster mass and 
the angle of the filament relative the line of the sight direction.  Fitting the constructed profile to the universal formula with 
constant amplitude and slope, we statistically determine the dynamical mass of each cluster and compare it with the previous estimates made 
by a conventional method.  The weak and strong points of the DME as well as its prospect for the measurements of the dynamical masses 
of the high-$z$ clusters are discussed.
\end{abstract}

\section{INTRODUCTION}\label{sec:intro}

Accurate measurement of the masses of the galaxy clusters is quite a narrow bottleneck to the success and completion of the cluster 
cosmology. Although a plethora of methodology has so far been developed to pass through this bottleneck, the required accuracy that would 
optimize the usage of the galaxy clusters as a probe of cosmology has yet to be achieved  \citep{cluster_cosmology}.  The conventional 
methodology obtains the masses of the galaxy clusters by finding and modeling their correlations with other observables such 
as the velocity dispersions of the member galaxies, the X-ray temperatures, the gravitational lensing signals, the Sunyaev-Zel'dovich (SZ) 
effect, the optical richness and so on.
The theoretical models for the correlations between the cluster masses and those observables, however, were often constructed by sacrificing 
the astrophysical complexities of real galaxy clusters which include deviation from hydrostatical/thermal equilibrium, incomplete 
relaxation of their dynamical states,  their non-spherical shapes, and existence of their substructures 
\citep[for a comprehensive review, see][]{scaling}

In fact, it is not only the low accuracy but the inconsistency that has to be hurdled in the measurements of the cluster masses.  
While the simplified assumptions about the correlations between the cluster masses and those observables yielded inaccurate measurements 
of the cluster masses, the variation of the degree of the simplification yielded the inconsistencies among the values of the cluster masses 
estimated by using different observables. It may be also responsible at least partly for the lower value of the linear power spectrum amplitude 
estimated by the abundance of the SZ clusters than the value measured by the Planck experiment of the cosmic microwave background 
radiation \citep{planck_xvi,planck_xx}. 

The previous attempts to deal with the reality were either statistical or resorting to the hydrodynamic simulations. The former approach 
statistically accounted for the intrinsic scatters of the correlations between the cluster masses and the observables, which ameliorated 
the accuracy but degraded the precision in the mass measurements of the galaxy clusters \citep[e.g.,][and references therein]{AH10}.  
The latter approach based on the hydrodynamic 
simulations made it possible to incorporate the astrophysical complexities into the models for the clusters but undermined the power of cluster 
cosmology since the detailed prescriptions of the baryon physics required to run hydrodynamic simulations are deeply cosmology-dependent
\citep[e.g.,][and references therein]{fin-etal10,stanek-etal10,pla-etal13,wu-etal15,tru-etal16}.

The algorithm recently developed by \citet{falco-etal14} has cleared a path through the above generic difficulties toward an independent 
measurement of the cluster masses.  It estimates the dynamic mass of a galaxy cluster by using the mass dependence of the recession  
velocity profile of the neighbor galaxies located in its bound zone.  Finding an empirical formula for the recession velocity profile from a 
N-body simulation and noting its universal behavior, \citet{falco-etal14} suggested that their algorithm should be particularly useful 
for those dynamically young unrelaxed clusters in the middle of merging process out of thermal equilibrium, which were difficult to 
deal with in the previous approaches.  \citet{falco-etal14} tested their algorithm against the Coma cluster to find a fairly good accord of 
their estimate with the previous measurements. From here on, we call this algorithm the Dynamic Mass Estimator (DME).

In the subsequent works, the DME has been further improved and refined. \citet{lee-etal15a} who attempted to estimate the dynamic mass 
of the Virgo cluster by using the DME algorithm pointed out that the original DME identifies a bound-zone filament in a somewhat haphazard 
way and suggested that for the identification of a true bound-zone filament it should be first examined if the recession velocities of the filament 
galaxies deviate from the Hubble flow.

\citet{lee16} have refined the analytic formula for the recession velocity profile of the filament galaxies by narrowing down 
the slopes and amplitudes that characterize the formula with the help of a higher resolution numerical simulation 
and showed that the universality of the formula for the recession velocity profile is valid only in the limited redshift range of $z\le 0.2$.
Although the refined DME algorithm pulled it off to estimate the dynamical mass of the nearest Virgo cluster \citep{lee-etal15a}, 
its power in the competition with the other conventional estimators has yet to be convincingly verified.  It is  essentially important 
to quantitatively explore with larger datasets how well it works in practice and what its success rate is, which we attempt to carry out in this 
Paper.
 
The contents of the upcoming Sections are summarized in the following. In Section \ref{sec:review} the DME algorithm is concisely reviewed. 
In Section \ref{sec:fil} we present the physical analyses of the group/cluster catalog from a large galaxy survey to identify the bound-zone 
filaments and to construct the recession velocity profiles along the filaments. In Section \ref{sec:mass} we present the dynamical mass 
estimates of the target clusters made by applying the DME to their bound-zone filaments and comparison of our results with the conventional 
estimates.  In Section \ref{sec:sum} the summary of the final results and the discussion on the future prospect for the application of the 
DME to the high-$z$ clusters are presented. 

\section{REVIEW OF THE DME ALGORITHM}\label{sec:review}

Consider a galaxy located in the bound zone around a massive cluster, where the separation distance between the galaxy and 
the cluster is large enough for the galaxy not to fall into the potential well of the cluster but also small enough for it to develop a 
non-negligible peculiar velocity.  The recession velocity of the bound-zone galaxy from the cluster should be lower than the Hubble speed 
but will gradually approach to it with the increment of the separation distance. 
The more massive the cluster is, the less rapidly the recession velocity of the bound-zone galaxy will change with the separation distance. 
Hence, the profile of the recession velocity of the bound-zone galaxy should be a powerful indicator of the cluster mass.

By using a N-body experiment, \citet{falco-etal14} have shown that the following formula provides a good approximation to the profile of the mean recession velocity of the bound-zone galaxies around a cluster with virial radius of $r_{v}$:
 \begin{equation}
 \label{eqn:vr}
 v_{r}(r) = H(z)r - A\,V_{c}\left(\frac{r}{r_{v}}\right)^{-n}\, ,
 \end{equation}
where $v_{r}(r)$ is the radial component of the recession velocity of a bound-zone galaxy at a distance $r$ from the cluster center, 
$V_{c}$ is the circular velocity at $r_{v}$, and $H(z)$ is the Hubble parameter. This formula has two free parameters, $A$ and $n$, that  
represent the amplitude and the slope of the profile, respectively. 

In the right-hand side (RHS) of Equation (\ref{eqn:vr}) the first term corresponds to the Hubble speed while the second term represents 
the peculiar velocity that depends on the cluster mass  $M_{v}$ through $r_{v}$ as $M_{v}=4\pi\,\Delta_{c}\,r^{3}_{v}/3$ where 
$\Delta_{c}\approx 100\rho_{c}$ and $\rho_{c}$ is the critical density of the Universe.  The two parameters $A$ and $n$ were 
determined to be $A=0.8\pm 0.2$ and $n=0.42\pm 0.16$ and claimed to be independent of the redshifts as well as of the key cosmological 
parameters by \citet{falco-etal14}. 

The subsequent analysis of \citet{lee16} based on the Millennium II simulations \citep{mill2} has confirmed that this assumption of 
\citet{falco-etal14} is valid but that the values of $A$ and $n$ are universal only in the limited range of $z\le 0.2$. 
\citet{lee-etal15b} demonstrated  a weak dependence of $A$ and $n$ on the mass scales with the help of the MultiDark  Planck simulations 
\citep{MDPL}  and determined their values with high precision as $A=0.88\pm 0.02$ and $n=0.43\pm 0.01$ on the cluster mass scale of 
$10^{14}\,h^{-1}M_{\odot}$.

If $v_{r}(r)$ and $r$ were measurable from observations, then the mass of a galaxy cluster would be readily estimated by putting the 
the observed profile $v_{r} $ to Equation (\ref{eqn:vr}). 
This principle, however, cannot be put into practice since we are not capable of measuring the profiles $v_{r}(r)$ and $r$ directly from 
observations.  
\citet{falco-etal14} put forth a clever idea to overcome this difficulty.  Suppose that some of the bound-zone galaxies around a cluster constitute 
a thin straight filament and that the filament is inclined at an angle of $\beta$ with the line-of-sight direction of the cluster. Their recession 
velocities, $v_{r}(r)$, and positions relative to the cluster center, $r$, can be expressed as $cz/\cos\beta$ and $r_{2d}/\sin\beta$, respectively, 
where $z$ denotes the relative redshift of the bound-zone galaxy from the cluster center and  $r_{2d}$ is the projected value of $r$ onto the 
plane of the sky perpendicular to the line-of-sight direction to the cluster, both of which are all directly observable. 
Rewriting Equation (\ref{eqn:vr}) in terms of these readily measurable quantities, we have 
 \begin{equation}
 \label{eqn:vr2d}
 \frac{cz(r_{2d},\beta,M_{v})}{\cos\beta} = 
\left[ H\frac{r_{2d}}{\sin\beta} - A\,V_{v}\left(\frac{r_{2d}}{\sin\beta\,r_{v}} \right)^{-n}\right]\, . 
 \end{equation}
The trade-off for expressing the profile in terms of the directly observables is having one more unknown quantity, $\beta$, in addition to 
$M_{v}$, which cannot help but degrade the precision in the measurements of $M_{v}$.  

The application of the DME to the target clusters will proceed as follows. 
Detect a thin straight filamentary structure of the galaxies in the bound-zone regions around a target cluster. Calculate the redshift difference 
between the cluster and each filament galaxy as well as the separation distance between them in the plane of the sky 
perpendicular to the line of sight direction toward the cluster to construct the recession velocity profile along the filament.  
Adjusting the observed profile to Equation (\ref{eqn:vr2d}) to find the best-fit values of the mass of the target cluster as well as 
the inclination angle of the bound-zone filament. In the following Section, we will apply this DME to the galaxy clusters in the local Universe. 

\section{APPLICATION OF THE DME TO THE SLOAN CLUSTERS}

\subsection{Detection of the Bound-Zone Filaments Around the SDSS Clusters}\label{sec:fil}
 
\citet{tempel-etal14} identified the groups of the galaxies in a flux-limited spectroscopic dataset of the galaxies from the Sloan Digital Sky 
Survey \citep[][hereafter, SDSS DR10]{sdssdr10} with the help of the modified Friends-of-Friends (FoF) group finder and 
compiled a catalog that contains spectroscopic information on the member galaxies belonging to each group as well as on the field galaxies. 
\citet{tempel-etal14} also provided information about $M_{nfw}$ and $N_{hern}$ of each cluster in the catalog where $M_{nfw}$ and 
$M_{hern}$ denote two different dynamical masses both of which were estimated from the radial velocity dispersions of the member galaxies. 
The difference between the two masses lies in the shape of the matter density profile of a cluster.   
The former $M_{nfw}$ was estimated under the assumption that the density profile is well approximated by the 
Navarro-Frenk-White (NFW) formula \citep{nfw96,nfw97}, while for the latter $M_{hern}$ the Hernquist profile \citep{hern90} was used. 
For a detailed description of the modified FoF finder and the catalog of the galaxy groups, see \citet{tempel-etal14}.

From the group catalog of \citet{tempel-etal14},  we select $29$ massive groups as the target clusters by the criterion that the number of the 
member galaxies should equal or exceed $100$, expecting that the gravitational influences of those massive clusters should be strong enough 
to be readily detectable in their bound zones.  Those groups with less than $100$ member galaxies are excluded to reduce the statistical noises 
in the mass measurement with the DME.  The more massive a cluster is, the longer filament it tends to have in its bound-zone. 
The longer bound-zone filament composed of a larger number of the neighbor galaxies suffers less from statistical noises.

We also select $533256$ galaxies as the sample galaxies from the spectroscopic dataset of the SDSS DR10 
by the criterion that the galaxies are either field or members of the low-mass groups with $10$ or less members.  The reason for excluding the 
galaxies belonging to the groups with more than $10$ members is as follows. In Equation (\ref{eqn:vr2d}), it is implicitly assumed that the 
most dominant gravitational influence on a galaxy with recession velocity $v_{r}(r)$ is from the cluster with mass $M_{v}$. If a galaxy in the 
neighbor region around a target cluster is a member of a group with more than $10$ members, then the gravitational effect of the other 
members belonging to the same host group on the galaxy may be comparable to that of the cluster and thus its peculiar velocity would no 
longer be well approximated by the second term in the RHS of Equation (\ref{eqn:vr2d}).
 
Now, to identify a bound-zone filament from the spatial distribution of the sample galaxies around each target cluster, we follow the prescriptions 
of \citet{falco-etal14}. First, we determine the relative redshifts, $z$, and the projected distances, $r_{2d}$, of the sample galaxies from the 
center of each target cluster. 
The former is obtained by taking the difference between the redshifts of the target cluster and its sample galaxies, while the latter is 
measured in the plane of the sky perpendicular to the line of sight direction to the cluster from information on their equatorial coordinates. 
Around each of the six target clusters, the sample galaxies which satisfy the conditions of $r_{2d}\le  20\,h^{-1}$Mpc and 
$\vert(cz)/H\vert\le 40\,h^{-1}$Mpc are selected as the neighbor galaxies, where $c$ is the speed of light. 
Dividing the ranges of $r_{2d}$ and $l_{z}\equiv (cz)/H$ into $4$ and $20$ bins, respectively, we pixelate the configuration space 
spanned by $r_{2d}$ and $l_{z}$ around each target cluster into $80$ squares each of which has an area of $dr_{2d}\,dl_{z}=16$.  

To compute the number density of the neighbor galaxies at each pixel, 
we also split the plane of the sky around each cluster into $8$ wedges according to the polar angles $\theta$ 
defined as $\theta = \tan^{-1}\left(x/y\right)$ in the range of $[0,2\pi)$ where $(x,y)$ is the Cartesian coordinates of a two dimensional 
position vector of the pixel center in the plane of the sky from a target cluster, satisfying the condition of $r_{2d}=\sqrt{x^{2}+y^{2}}$.  
The $k$-th wedge corresponds to the $\theta$-interval of $[(k-1)\pi/4,\ k\pi/4)$ where the integer $k$ varies from $1$ to $8$. 
Grouping the pixels in the $r_{2d}$-$l_{z}$ plane by the polar angles of the position vectors of the centers of the pixels, we end up having 
eight different realizations from the eight wedges for the number density of the neighbor galaxies at each pixel. 

Let $n_{ij}^{k}$ be the number density of the neighbor galaxies belonging to the $ij$-th pixel 
(i.e., the $i$-th bin of of $r_{2d}$ and the $j$-th bin of $l_{z}$) from the $k$-th wedge.
The dimensionless density contrast, $\delta_{ij}^{k}$, can be calculated as $\delta_{ij}^{k}\equiv (n_{ij}^{k}-\bar{n}_{ij}^{k})/\bar{n}_{ij}^{k}$. 
Here, we evaluate the mean number density $\bar{n}_{ij}^{k}$ by taking the ensemble average over the number densities at the same $ij$-th 
pixel but from the five different wedges, excluding the realizations from the $k$-th wedge and its two adjacent wedges. 
Then, we select only those pixels that meet the condition of $\delta_{ij}^{k}\ge 3$ as the candidiate overdense sites where the bound-zone 
filaments may be found, as done in \citet{falco-etal14}.  

\begin{figure}
\begin{center}
\plotone{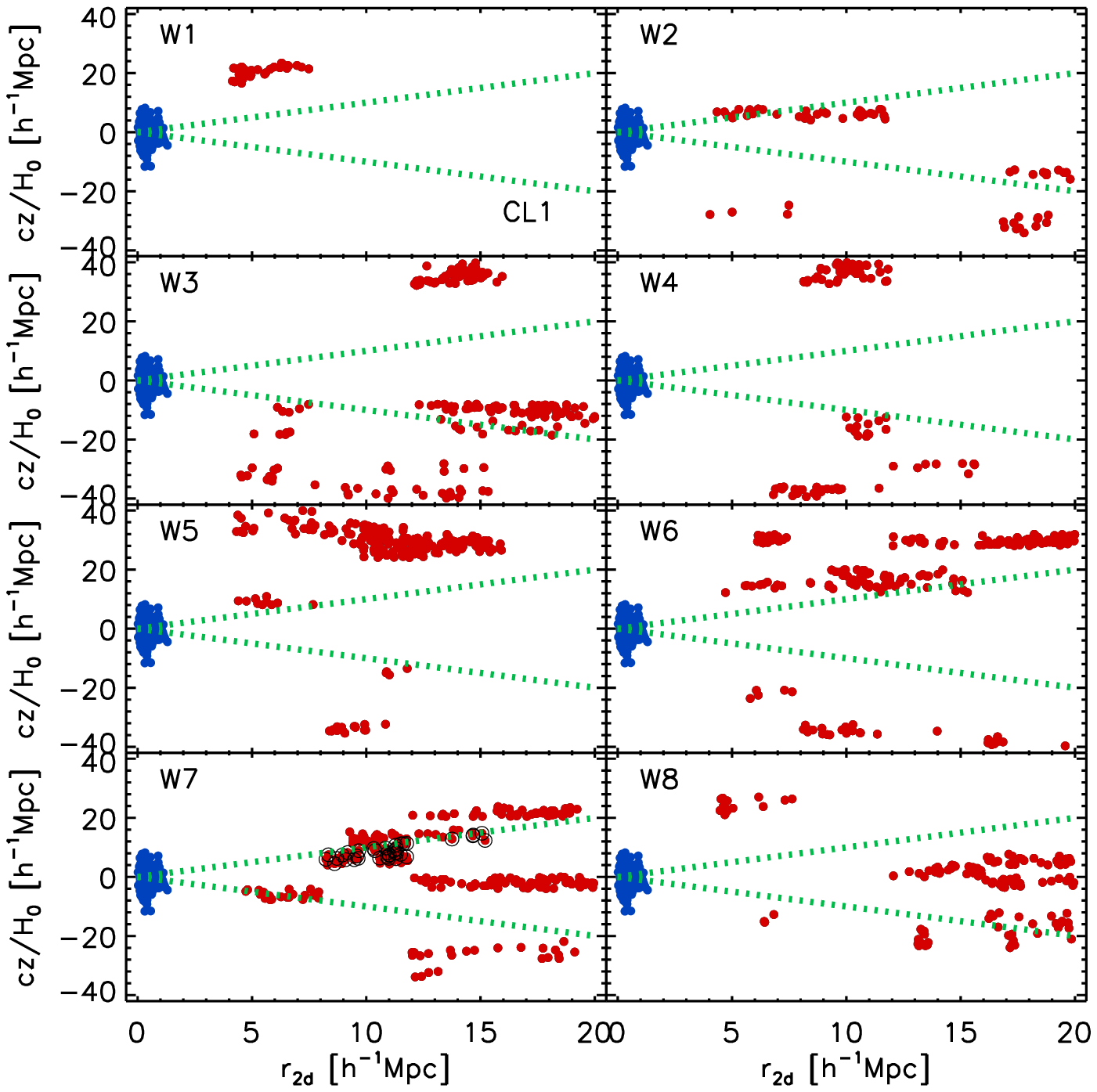}
\caption{Recession velocities along the line of sight directions versus the projected distances in the plane of the sky 
from the eight different wedges (W1-W8). In each panel, the filled blue circles correspond to the configurations of the member 
galaxies of  one of the six target clusters (CL1) while the filled red circles represent those of the neighbor galaxies belonging to the 
candidate overdense pixels, and the open black circles represent the sloping straight line-like filaments steeper than the Hubble flow 
displayed as dashed green lines.}
\label{fig:cl1}
\end{center}
\end{figure}

Eight panels of Figure \ref{fig:cl1} depict the distributions of the neighbor galaxies belonging to the candidate overdense pixels with 
$\delta_{ij}^{k}\ge 3$ (red dots) from the eight wedges (W1-W8) in the $r_{2d}$-$l_{z}$ configuration space around one of the 
$29$ target clusters. 
In each panel, the blue dots represent the configurations of the member galaxies of the target cluster (dubbed CL1).
In the original procedure described by \citet{falco-etal14}, a bound-zone filament was identified in the distribution of the overdense pixels as a 
sloping straight line which exhibits a monotonic increment of $\vert l_{z}\vert$ with $r_{2d}$. 
As mentioned in Section \ref{sec:review}, \citet{lee-etal15a} suggested that a bound-zone filament should be identified not just as a sloping 
straight line but satisfy an additional condition which is in fact essential to the success of the DME. 
If the neighbor galaxies belonging to a bound-zone filament is under the dominant gravitational influence of a target cluster, then their 
recession velocities should be lower than the Hubble speed at small distances but gradually approach to it as the distances increase 
\citep[see][]{kim-etal16}. 
In other words, a bound-zone filament should appear as a sloping straight line steeper than the straight line of $\vert l_{z}\vert=r_{2d}$ 
which is plotted as green dotted line in each panel of Figure \ref{fig:cl1}. 

A shrewd reader might think that this condition is too stringent since $r_{2d}$ is not a real three dimensional distance, $r$, between a target 
cluster and its neighbor galaxies but only a two dimensional distance projected onto the plane of the sky. Yet, without having any information on 
$r$, it is the most conservative and secure condition required to guarantee the validity of the DME.  We look for such a sloping straight line 
steeper than the green solid line in the distributions of the red dots and identify one in the realization from the seventh wedge (W7), which is 
shown as the open black circles overlapped with the red closed circles in the bottom left panel of Figure \ref{fig:cl1}.

\begin{figure}
\begin{center}
\plotone{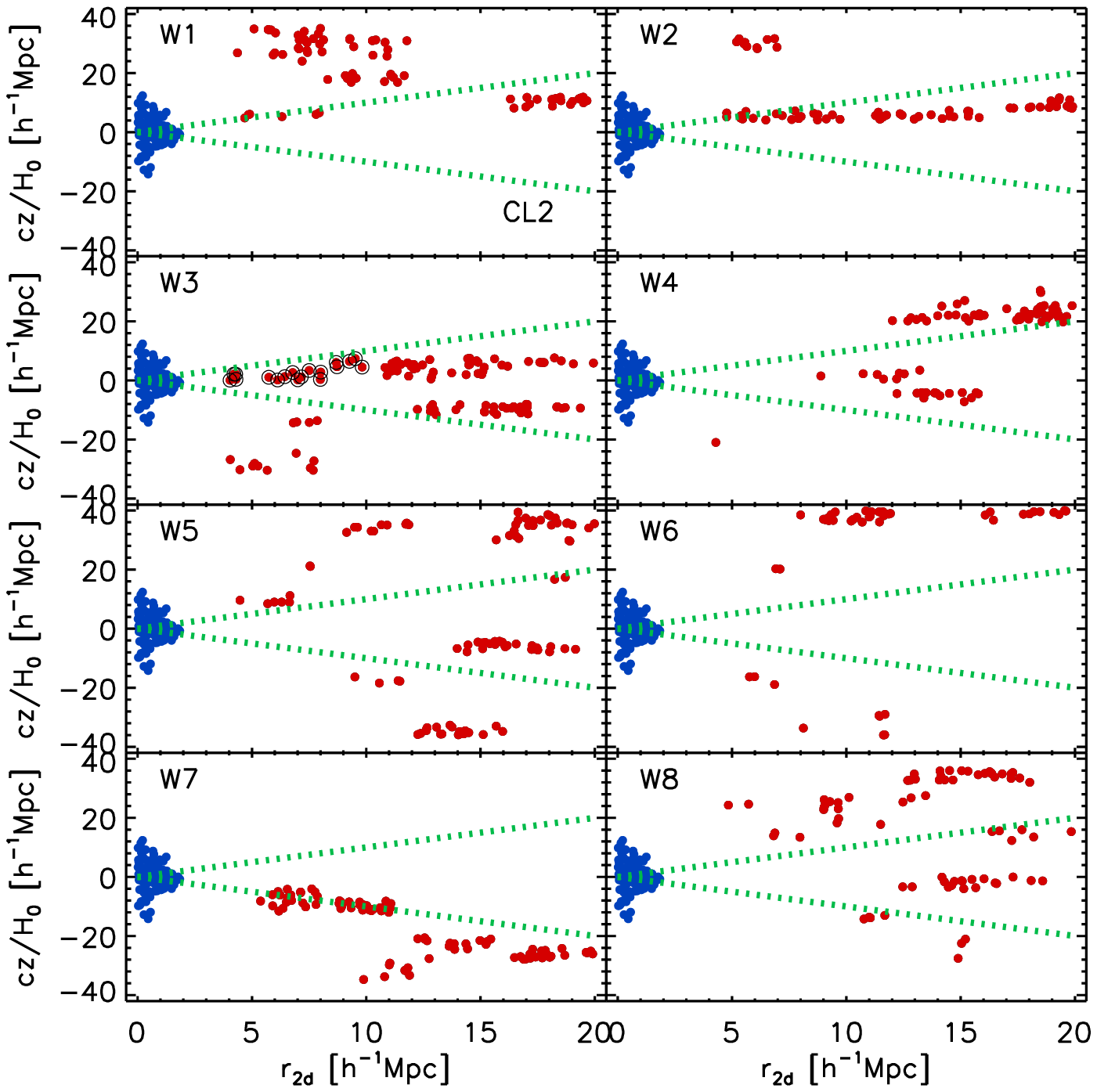}
\caption{Same as Figure \ref{fig:cl1} but with for the CL2.}
\label{fig:cl2}
\end{center}
\end{figure}

\begin{figure}
\begin{center}
\plotone{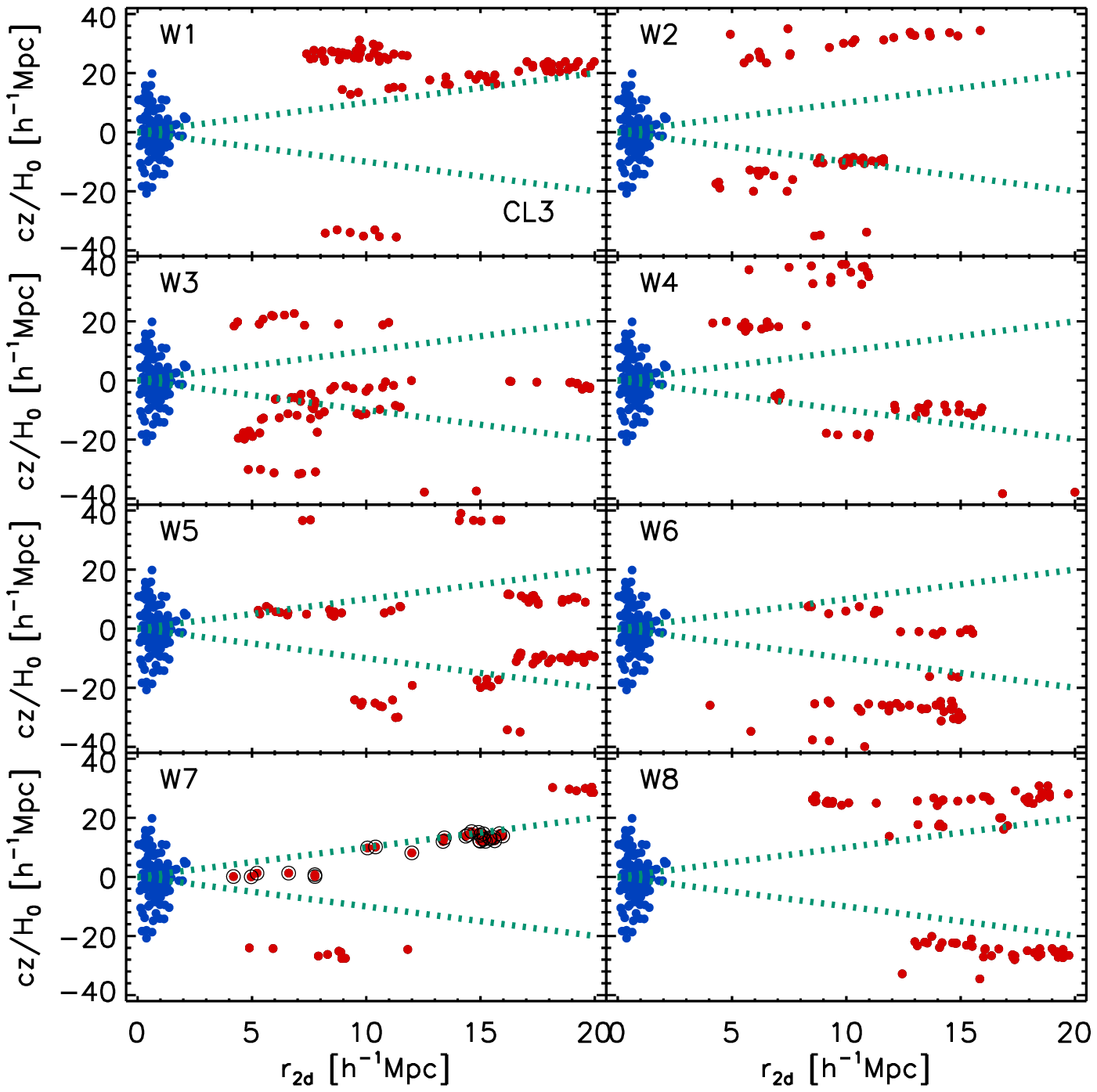}
\caption{Same as Figure \ref{fig:cl1} but with for the CL3.}
\label{fig:cl3}
\end{center}
\end{figure}

\begin{figure}
\begin{center}
\plotone{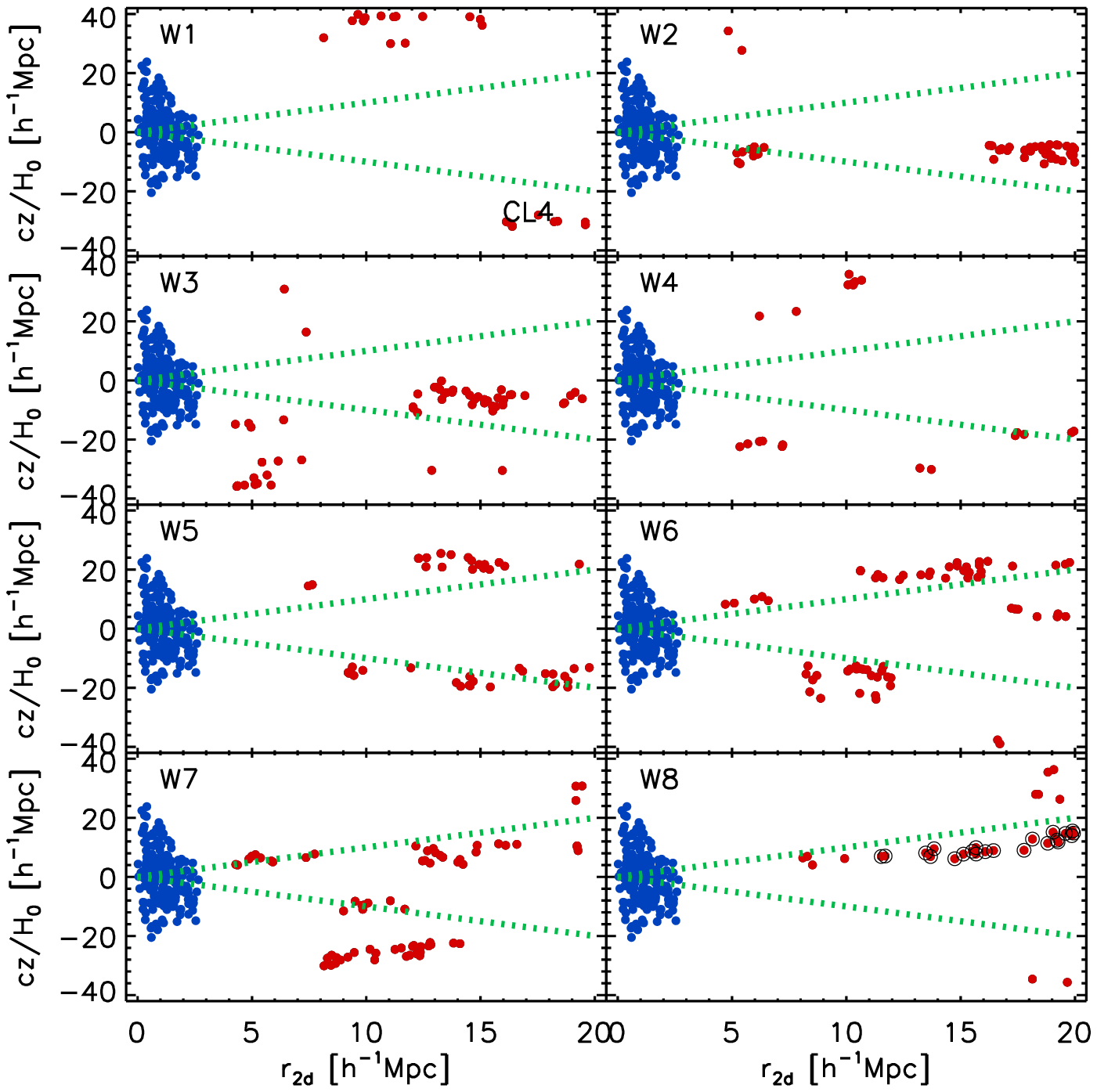}
\caption{Same as Figure \ref{fig:cl1} but with for the CL4.}
\label{fig:cl4}
\end{center}
\end{figure}

\begin{figure}
\begin{center}
\plotone{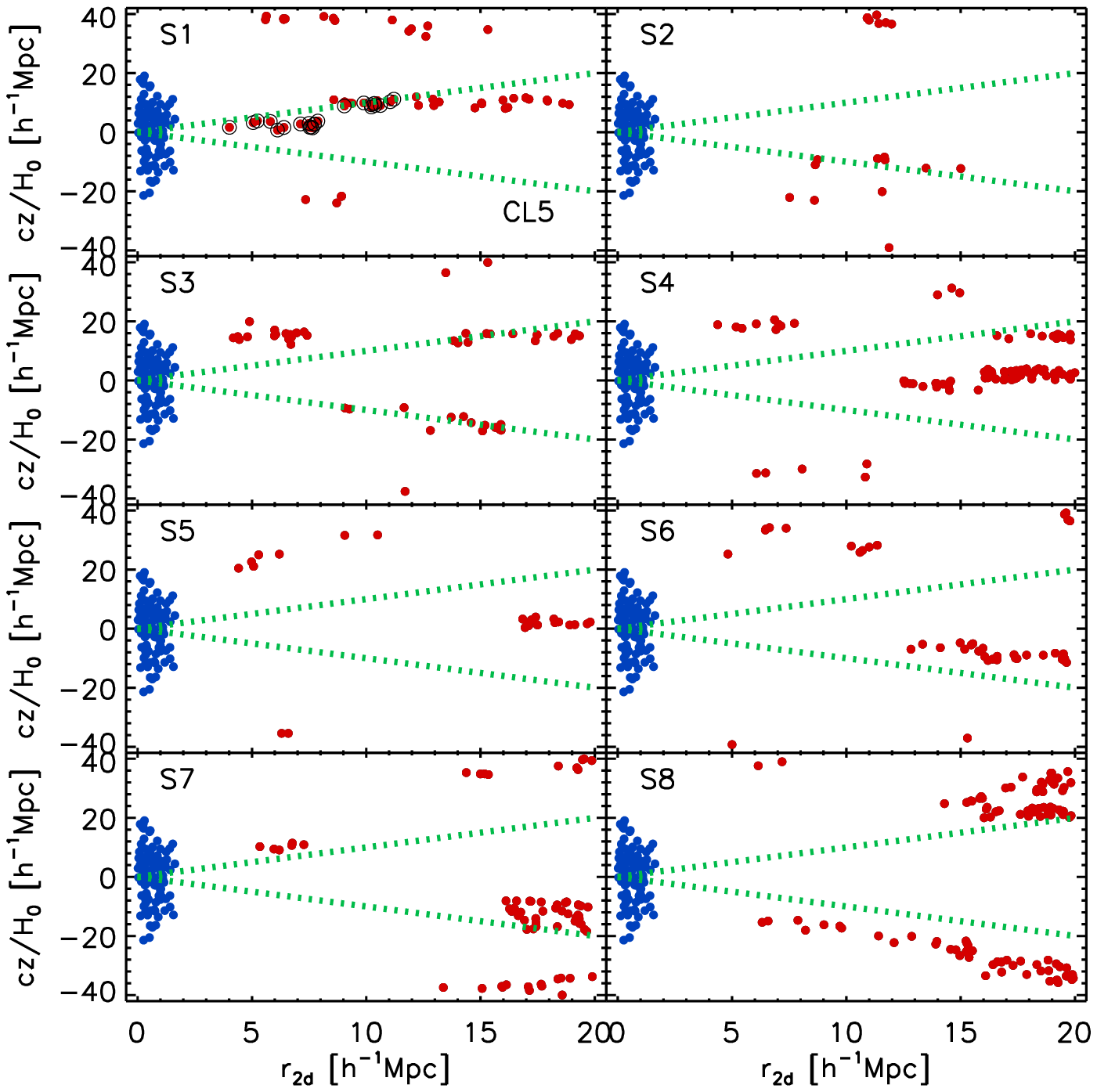}
\caption{Same as Figure \ref{fig:cl1} but with for the CL5.} 
\label{fig:cl5}
\end{center}
\end{figure}

\begin{figure}
\begin{center}
\plotone{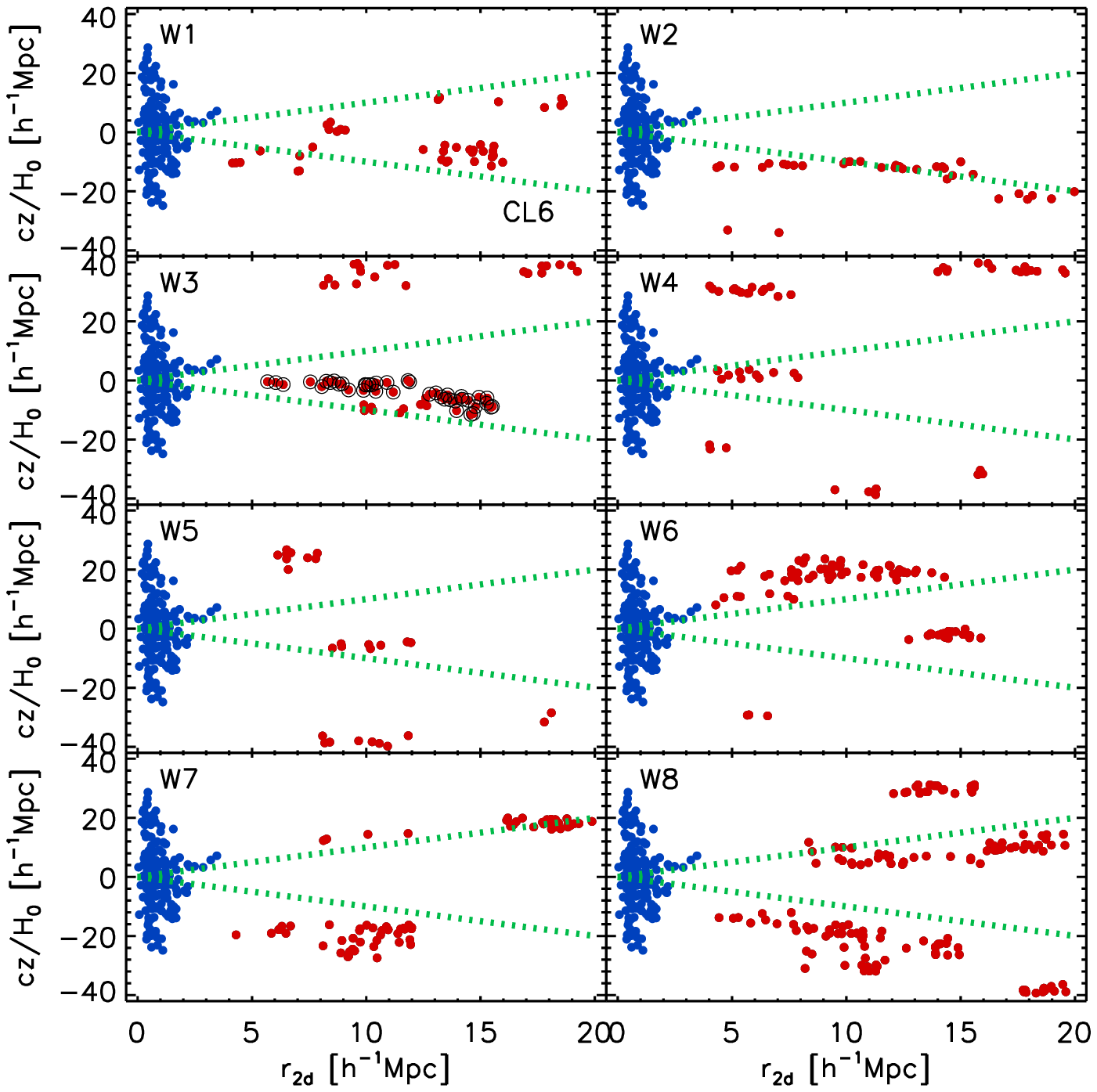}
\caption{Same as Figure \ref{fig:cl1} but with for the CL6.} 
\label{fig:cl6}
\end{center}
\end{figure}

\begin{figure}
\begin{center}
\plotone{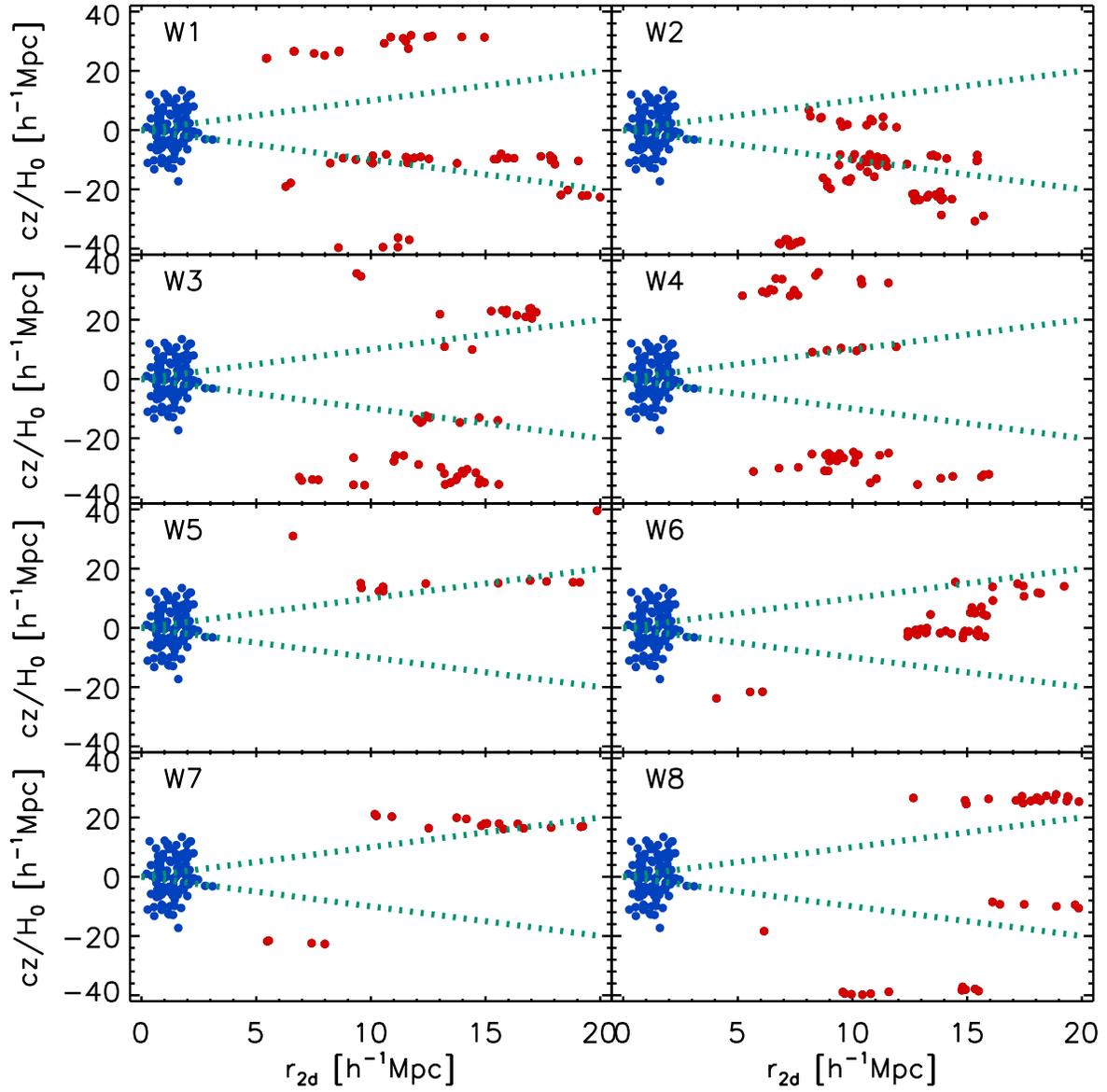}
\caption{Same as Figure \ref{fig:cl1} but with for the case of a Sloan cluster around which no bound-zone 
filament is detected.}
\label{fig:cl7}
\end{center}
\end{figure}

We followed the same procedures to find the bound-zone filaments of the $29$ target clusters and found that only six targets have such 
thin straight filamentary structures in their bound-zones. Figures \ref{fig:cl2}-\ref{fig:cl6} display the same as Figure \ref{fig:cl1} but for 
the other five clusters (CL2-CL6).  Table \ref{tab:cl} lists the identification number, (Group ID), spectroscopic redshifts, equatorial coordinates, 
the numbers of the member galaxies ($N_{m}$), the numbers of the neighbor galaxies belonging to the bound-zone filaments ($N_{g}$), of the 
six target clusters. Regarding the other $23$ target clusters, we fail to find the bound-zone filaments from the configurations of the neighbor galaxies 
in the $r_{2d}$-$l_{z}$ space. Figure \ref{fig:cl7} shows one example of a target cluster in the bound zone of which no thin straight 
filamentary structure is found in any of the eight wedges. 

\subsection{Estimates of the Dynamic Masses of Six Clusters with DME}\label{sec:mass}

For each of the six clusters in the bound-zone of which a thin straight filamentary structure is detected in Section \ref{sec:fil}, we fit 
the observational results to Equation (\ref{eqn:vr2d}) to simultaneously find the best-fit values of  $m_{v}\equiv \log M_{v}/(h^{-1}M_{\odot})$ 
and $\beta$ that maximizes the likelihood distribution of $p\left[-\chi^{2}(m_{v},\beta)/2\right]$ where $\chi^{2}(m_{v},\beta)$ is given as 
 \begin{equation}
 \label{eqn:chi2}
\chi^{2}\equiv 
\sum_{i=1}^{N_{f}}\left\{\frac{l_{z,i}}{\cos\beta} - \left[\frac{r_{2d,i}}{\sin\beta} - 
A\,\frac{V_{v}}{H}\left(\frac{r_{2d,i}}{\sin\beta\,r_{v}} \right)^{-n}\right]\right\}\frac{1}{\sigma^{2}_{i}}\, . 
 \end{equation}
 where $l_{z,i}$ and $r_{2d,i}$ denote the observed values of $l_{z}$ and $r_{2d}$ of the $i$-th galaxy belonging to the bound-zone  filament.  As done in \citet{lee-etal15a}, the one standard deviation errors $\sigma_{i}$ are all set at unity, given that the uncertainties 
associated with the measurements of $l_{z}$ including the ones associated the identification of the bound-zone filaments by the eyes 
 are unknown.  

\begin{figure}
\begin{center}
\plotone{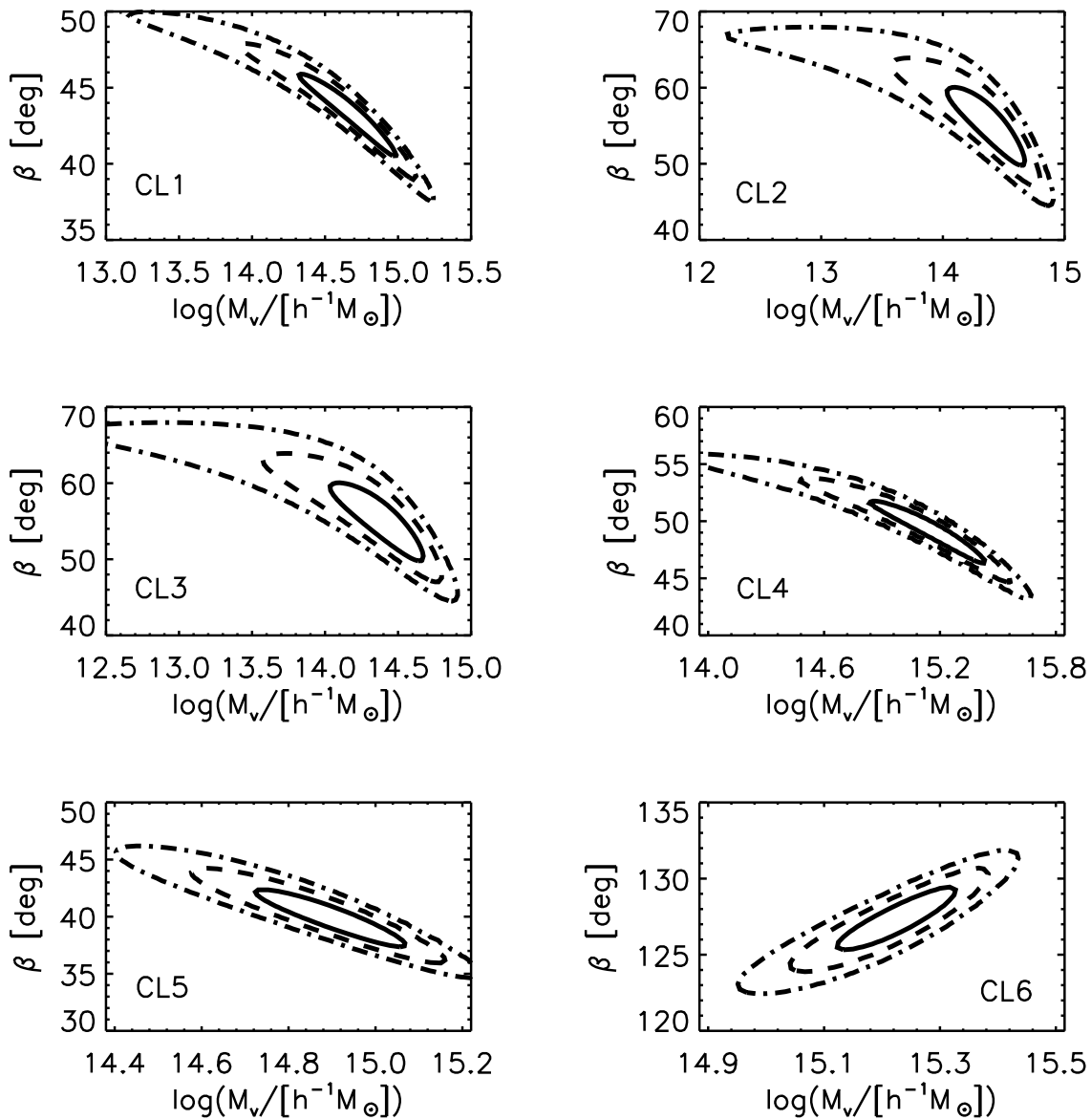}
\caption{$68\%,\ 95\%,\ 99\%$ contours of the likelihood distributions in the plane spanned by the logarithmic 
mass $M_{v}$ and the inclination angle $\beta$ for the six clusters around which the bound-zone filaments 
are detected.}
\label{fig:cont}
\end{center}
\end{figure}

\begin{figure}
\begin{center}
\plotone{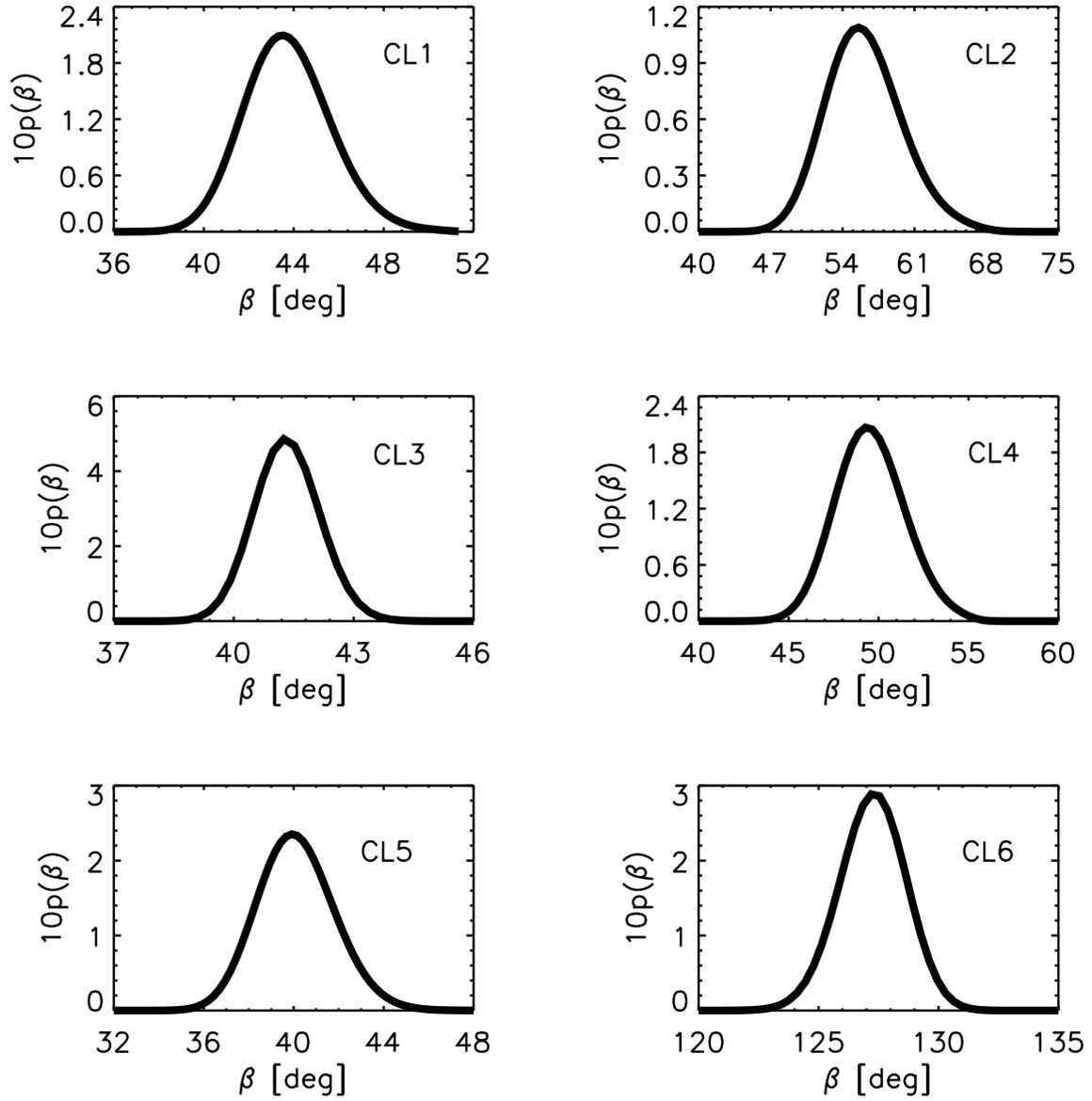}
\caption{Probability density functions of the inclination angles of the bound-zone filaments around the 
six clusters marginalized over the logarithmic masses. }
\label{fig:pbeta}
\end{center}
\end{figure}

Each panel of Figure \ref{fig:cont} displays the $68\%,\ 95\%,\ 99\%$ contours of the likelihood in the $m_{v}$-$\beta$ plane as the solid, 
dashed and dot-dashed lines, respectively, for each case of the six clusters.  As can be seen, for all cases of the six clusters, the $68\%$ 
contours are well localized.   
Marginalizing $p\left[-\chi^{2}(m_{v},\beta)/2\right]$ over $m_{v}$ as $p(\beta) = \int_{-\infty}^{\infty}\,p(m_{v},\beta)dm_{v}$,  we obtain the 
one-point probability density function, $p(\beta)$, for the six clusters, the results of which are shown in Figures \ref{fig:pbeta}.
As can be seen, for the case of the CL3 which has the thinnest bound-zone filament among the six,  the probability density function 
$p(\beta)$ has the narrowest shape. Whereas, for the cases of the CL2 and CL6 whose bound-zone filaments appear relatively thick in the 
$r_{2d}$-$l_{z}$ plane, a widely spreaded shape of $p(\beta)$ is noted. 

\begin{figure}
\begin{center}
\plotone{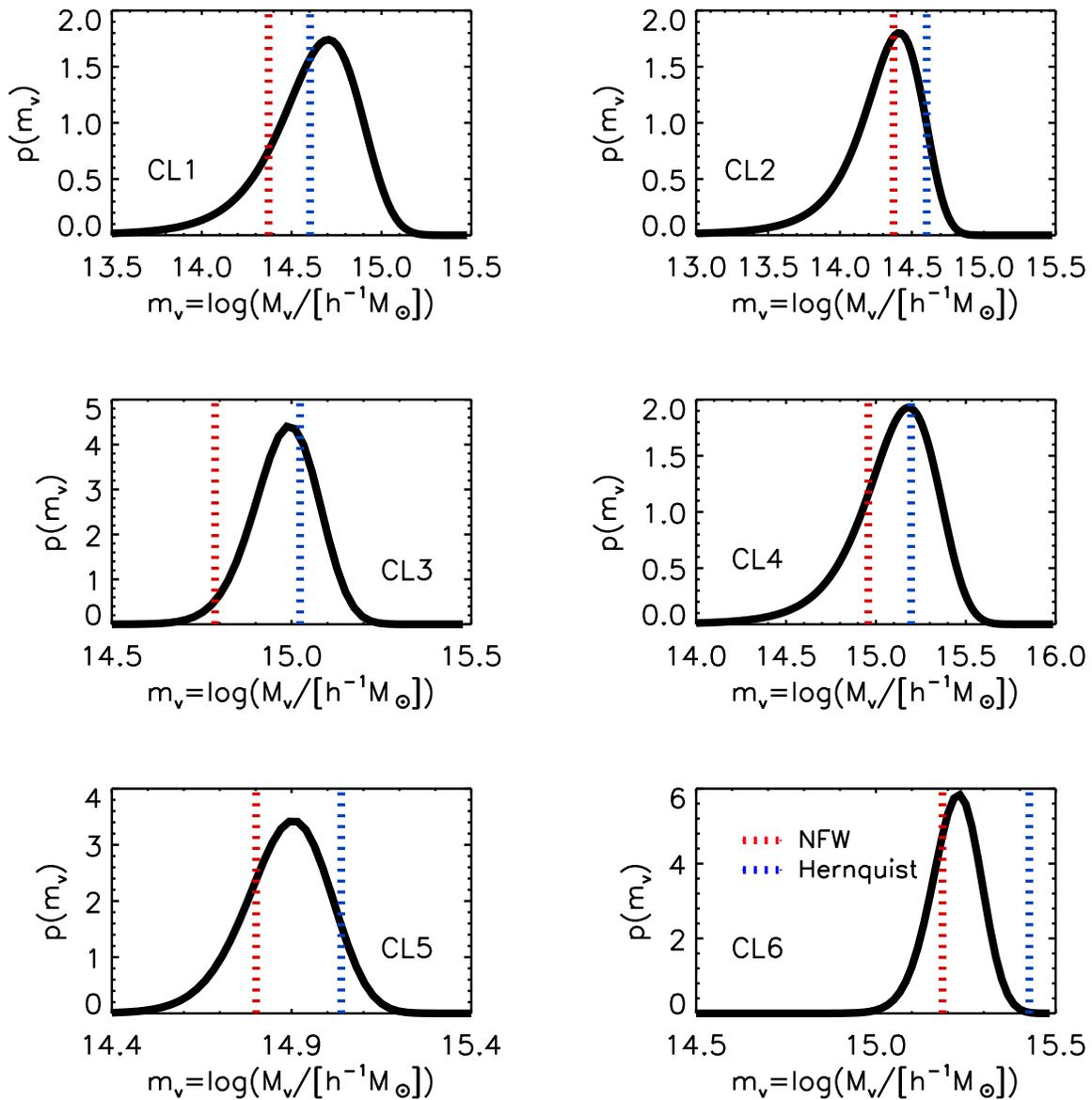}
\caption{Probability density functions of the dynamical masses of the clusters marginalized over the inclination 
angles of the bound-zone filaments.}
\label{fig:pmv}
\end{center}
\end{figure}

Marginalizing $p\left[-\chi^{2}(m_{v},\beta)/2\right]$ over $\beta$ yields the one-point probability density function, $p(m_{v})$, the results of 
which are shown in Figures \ref{fig:pmv}. 
As can be seen, the probability density distributions deviate from the Gaussian shape, asymmetric around the best-fit value at which 
$p(m_{v})$ reaches it maximum. For the case of the CL6 whose bound-zone filaments have the largest number of the neighbor galaxies, 
$N_{g}$, the shape of $p(m_{v})$ is the closest to the Gaussian distribution, which indicates that the asymmetric shape of $p(m_{v})$ 
is likely due to the small number statistics. 
The red and blue dotted lines in each panel of Figure \ref{fig:pmv} correspond to the two mass estimates of each of the six clusters made 
by \citet{tempel-etal14} with the conventional method based on the radial velocity dispersions of the cluster galaxies: 
$m_{nfw}\equiv \log M_{nfw}/(h^{-1}M_{\odot})$ and $m_{hern}\equiv \log M_{hern}/(h^{-1}M_{\odot})$.  As can be seen, although 
our best-fit values of $m_{v}$ do not show significant difference from $m_{nfw}$ and $m_{hern}$, the amount and trend of the difference 
changes from cluster to cluster. 
For the case of the CL1, our best-fit value of $m_{v}$ exceeds both of $m_{nfw}$ and $m_{hern}$. For the case of the CL2,  
our result agrees well with $m_{nfw}$. For the cases of the CL3 and CL4,  our estimates coincides with $m_{hern}$.  
For the other two cases of CL5 and CL6, our best-fit values of $m_{v}$ lie between $m_{nfw}$ and $m_{hern}$.

\section{DISCUSSION AND CONCLUSION}\label{sec:sum}

The DME algorithm developed by \citet{falco-etal14} estimates the mass of a galaxy cluster from the profile of the recession velocities of the 
neighbor galaxies constituting a thin straight filament in its bound zone. Therefore, the application of the DME is inherently limited to the galaxy 
clusters in the bound-zones of which thin straight line-like filaments exist.  In the current analysis, only six out of the $29$ clusters composed of 
$100$ or more member galaxies in the SDSS group catalog are found to have such thin straight filaments in their bound zones, which implies 
that the success rate of DME should be around $20\%$. 

The other downside of the DME is the somewhat casual way in which thin straight filaments are identified in the bound zones.
From the distributions of the neighbor galaxies in the configuration space spanned by their redshifts and projected distances, 
the sloping straight line-like structures steeper than the Hubble flow had to be sought after by the eyes as the bound-zone filaments
\citep{falco-etal14}. 
It would be quite desirable to construct a more formal deliberate routine for the detection of a bound-zone filament from the spatial 
distributions of the neighbor galaxies. 

Nevertheless, the above downsides of the DME do not overshadow its distinct advantage over the other conventional mass estimators. 
Since the DME requires no simplified assumptions about the dynamical and/or thermal states nor about the shapes and profiles of the 
clusters, it can be applied even to those clusters which are in the middle of merging process, having very disturbed shapes with very low 
X-ray/SZ emissions. The dynamic masses of the six Sloan clusters estimated by the DME in the current analysis are found to be 
not substantially different from the previous estimates made by using the conventional methods based on the radial velocity dispersions 
of the cluster galaxies under the assumptions that the galaxy clusters are well relaxed having spherically symmetric shapes.  Finding that the 
amount of the difference between our estimate and the previous ones change from cluster to cluster, we suggest that the DME should be also 
useful to examine the deviation of the dynamical/thermal states of the clusters from the equilibrium and the asymmetry of their true density 
profiles as well.

The usefulness of the DME confirmed in the current analysis leads us to expect that the DME may be an optimal algorithm for the 
measurements of the dynamic masses of the high-$z$ clusters.  It has been the gravitational lensing and/or the SZ effects that have been 
almost exclusively employed to estimate the masses of the high-$z$ clusters not only because the velocity dispersions of the cluster galaxies at 
high redshifts are difficult to determine with high accuracy but also because the high-$z$ clusters are often dynamically young through 
merger events with abundant substructures, for which cases the previous dynamic mass estimators are likely to fail.     
Moreover, modified gravity (MG) models generically predict the dynamical masses of the galaxy clusters to be higher than the masses 
estimated by using the gravitational lensing effects \citep[e.g.,][]{schmidt-etal10}. 
Measuring the dynamic masses of the high-$z$ galaxy clusters with the DME and comparing them with the lensing counterparts would allow us 
to efficiently test the gravity, which is the direction of our future work.

\acknowledgements

This research was supported by Basic Science Research Program through the National Research Foundation of Korea(NRF) funded by the 
Ministry of Education(2016R1D1A1A09918491). It was also partially supported by a research grant from the NRF to the Center for Galaxy 
Evolution Research (NO. 2010-0027910).

\clearpage
\clearpage
\clearpage
\clearpage
\clearpage
\clearpage
\clearpage
\clearpage
\clearpage
\clearpage
\clearpage
\begin{deluxetable}{ccccccc}
\tablewidth{0pt}
\setlength{\tabcolsep}{5mm}
\tablecaption{Numbers of the member galaxies of the six clusters and their bound-zone filaments}
\tablehead{ & Group ID & redshift & RA & DEC & $N_{m}$ & $N_{g}$ \\ & & & (deg) & (deg) & }
\startdata
CL1 & $175$ & $0.02$ & $181.1$ & $20.4$ & $139$  & $36$  \\
CL2 & $1701$ & $0.05$ & $169.1$ & $29.3$ & $113$  & $18$  \\
CL3 & $2111$ & $0.07$ & $190.3$ & $18.6$ & $120$  & $27$  \\
CL4 & $3070$ & $0.09$ & $239.5$ & $27.3$ & $212$  & $22$  \\
CL5 & $5278$ & $0.08$ & $184.4$ & $3.7$  & $106$  & $23$  \\
CL6 & $7045$ & $0.07$ & $230.7$ & $27.8$ & $161$  & $41$ \\
\enddata
\label{tab:cl}
\end{deluxetable}

\clearpage
\begin{thebibliography}{000}
\bibitem[Ahn et al.(2014)]{sdssdr10} 
Ahn, C.~P., Alexandroff, R., Allende Prieto, C., et al.\ 2014, \apjs, 211, 17 
\bibitem[Allen et al.(2011)]{cluster_cosmology} 
Allen, S.~W., Evrard, A.~E., \& Mantz, A.~B.\ 2011, \araa, 49, 409 
\bibitem[Andreon \& Hurn(2010)]{AH10} 
Andreon, S., \& Hurn, M.~A.\ 2010, \mnras, 404, 1922 
\bibitem[Boylan-Kolchin et al.(2009)]{mill2} 
Boylan-Kolchin, M., Springel, V., White, S.~D.~M., Jenkins, A., \& Lemson, G.\ 2009, \mnras, 398, 1150
\bibitem[Evrard et al.(2014)]{evrard-etal14} 
Evrard, A.~E., Arnault, P., Huterer, D., \& Farahi, A.\ 2014, \mnras, 441, 3562
\bibitem[Falco et al.(2014)]{falco-etal14} 
Falco, M., Hansen, S.~H., Wojtak, R., et al.\ 2014, \mnras, 442, 1887 
\bibitem[Finoguenov et al.(2010)]{fin-etal10} 
Finoguenov, A., Sanderson, A.~J.~R., Mohr, J.~J., Bialek, J.~J., \& Evrard, A.\ 2010, \aap, 509, A85 
\bibitem[Giodini et al.(2013)]{scaling} 
Giodini, S., Lovisari, L., Pointecouteau, E., et al.\ 2013, \ssr, 177, 247 
\bibitem[Hernquist(1990)]{hern90} 
Hernquist, L.\ 1990, \apj, 356, 359 
\bibitem[Kim et al.(2016)]{kim-etal16} 
Kim, S., Rey, S.-C., Bureau, M., et al.\ 2016, \apj, 833, 207
\bibitem[Klypin et al.(2016)]{MDPL} 
Klypin, A., Yepes, G., Gottl{\"o}ber, S., Prada, F., \& He{\ss}, S.\ 2016, \mnras, 457, 4340 
\bibitem[Lee et al.(2015a)]{lee-etal15a} 
Lee, J., Kim, S., \& Rey, S.-C.\ 2015a, \apj, 807, 122
\bibitem[Lee et al.(2015b)]{lee-etal15b} 
Lee, J., Kim, S., \& Rey, S.-C.\ 2015b, \apj, 815, 43
\bibitem[Lee(2016)]{lee16} 
Lee, J.\ 2016, \apj, 832, 123 
\bibitem[Lee \& Yepes(2016)]{LY16} 
Lee, J., \& Yepes, G.\ 2016, \apj, 832, 185
\bibitem[Lee \& Li(2016)]{LL16} 
Lee, J., \& Li, B.\ 2016, arXiv:1610.07268 
\bibitem[Navarro et al.(1996)]{nfw96} 
Navarro, J.~F., Frenk, C.~S., \& White, S.~D.~M.\ 1996, \apj, 462, 563 
\bibitem[Navarro et al.(1997)]{nfw97} 
Navarro, J.~F., Frenk, C.~S., \& White, S.~D.~M.\ 1997, \apj, 490, 493 
\bibitem[Planelles et al.(2013)]{pla-etal13} 
Planelles, S., Borgani, S., Dolag, K., et al.\ 2013, \mnras, 431, 1487 
\bibitem[Planck Collaboration et al. XVI(2014a)]{planck_xvi} 
Planck Collaboration, Ade, P.~A.~R., Aghanim, N., et al.\ 2014, \aap, 571, A16
\bibitem[Planck Collaboration et al. XX(2014b)]{planck_xx} 
Planck Collaboration, Ade, P.~A.~R., Aghanim, N., et al.\ 2014, \aap, 571, A20
\bibitem[Schmidt(2010)]{schmidt-etal10} 
Schmidt, F.\ 2010, \prd, 81, 103002
\bibitem[Stanek et al.(2010)]{stanek-etal10} 
Stanek, R., Rasia, E., Evrard, A.~E., Pearce, F., \& Gazzola, L.\ 2010, \apj, 715, 1508
\bibitem[Tempel et al.(2014)]{tempel-etal14} 
Tempel, E., Tamm, A., Gramann, M., et al.\ 2014, \aap, 566, A1 
\bibitem[Truong et al.(2016)]{tru-etal16} 
Truong, N., Rasia, E., Mazzotta, P., et al.\ 2016, arXiv:1607.00019 
\bibitem[Wu et al.(2015)]{wu-etal15} 
Wu, H.-Y., Evrard, A.~E., Hahn, O., et al.\ 2015, \mnras, 452, 1982
\end{thebibliography}
\end{document}